# First-principle investigation of electronic structure, magnetism and phase stability of Heusler-type Pt$_{2-x}$Mn$_{1+x}$Ga alloys

L. Feng[1,a)] E. K. Liu[2], W. X. Zhang[1], W. H. Wang[2], G. H. Wu[2]

[1]*Key Laboratory of Advanced Transducers and Intelligent Control System, Ministry of Education, Computational Condensed Matter Physics Laboratory, Department of Physics, Taiyuan University of Technology, Taiyuan 030024, People's Republic of China*

[2]*Beijing National Laboratory for Condensed Matter Physics, Institute of Physics, Chinese Academy of Sciences, Beijing 100190, People's Republic of China*

Abstract: The electronic structure, magnetism and phase stability of Pt$_{2-x}$Mn$_{1+x}$Ga(x=0, 0.25, 0.5, 0.75, 1) alloys are studied by first-principle calculations. The calculations reveal that a potential magnetic martensitic transformation can be expected in all the series. In addition, a large magnetic-field-induced strain is likely to appear in Pt$_{2-x}$Mn$_{1+x}$Ga(x=0, 0.25, 0.75, 1) alloys. The electronic structure calculations indicate that the tetragonal phase is stabilized upon the distortion because of the pseudogap formation at the Fermi Level. The magnetic structure is also investigated and the total magnetic moment of the tetragonal phase is a little larger than that of the cubic austenite phase in all the series.

1. Introduction

Since Heusler ferromagnetic shape memory alloys(FSMAs) have exhibited diverse functional properties, such as magnetic-field-induced shape memory effects(MFISME), magnetic-field-induced strains(MFIS), magnetoresistance(MR), Hall effect, magnetocaloric effects(MCE), and so on [1-5], great efforts have been made to develop new-type FSMAs. Up to now, many investigations on the materials with the martensitic transformation have been done based on the first-principles calculations. The martensitic transformation and magnetic properties obtained based on these calculations are in good agreement with the experimental results [6-13]. Thus, the first-principles calculations have become an effective method to search for new-type

---

a) Corresponding author.
   E-mail address: fenglin@tyut.edu.cn

FSMAs. Mn$_2$-based Heusler alloys, which possess small magnetic moment, high Curie temperature, large coercivity, large exchange bias and remarkable magnetocrystalline anisotropy, have attracted great attention in magnetic material field [12, 14-15]. Recently, in Mn$_2$PtGa, with a tetragonal Heusler structure at room temperature, a large zero field exchange bias is reported [15]. This tetragonal Heusler structure indicates a possibility of the occurrence of the martensitic transformation above the room temperature. On the other hand, it has been reported that there is martensitic transformation in Pt$_2$MnGa [16]. Thus, it is meaningful to investigate the martensitic transformation and magnetic properties of the intermediate composition Pt$_{2-x}$Mn$_{1+x}$Ga alloys to search for the new-type FSMAs. In this work, we investigate the phase stability, electronic structure and magnetic properties of Heusler-type Pt$_{2-x}$Mn$_{1+x}$Ga(x=0, 0.25, 0.5, 0.75, 1) alloys using the first-principles calculations. The calculated results reveal that a potential magnetic martensitic transformation with a large thermodynamic driving force and a moderate strain energy can be expected in this system. Besides, a large MFIS is likely to appear in Pt$_{2-x}$Mn$_{1+x}$Ga(x=0, 0.25, 0.75, 1) alloys.

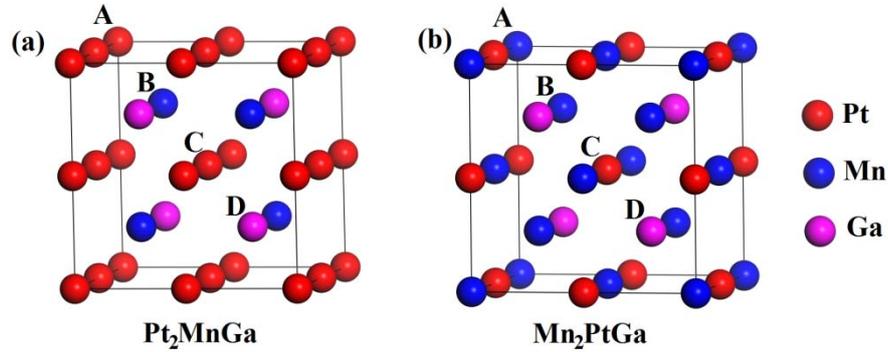

Fig. 1. The atomic configuration of stoichiometric Pt$_2$MnGa (a) and Mn$_2$PtGa (b).

2. Computational Details

In this study, the possibility of the martensitic transformation for cubic Heusler alloys is predicted based on the tetragonal distortion. That is, we simulate the phase transition by comparing the total energy of the ground state of cubic austenite and non-modulated tetragonal distorted phases with different axis ratio(c/a). In the calculations, the volume of unit cell is assumed to be constant during the tetragonal distortion. Here, it should be pointed out that, same to the case in many preceding work, the tetragonal distortion is made based on the Heusler-type conventional cell with 16 atoms. The spin-polarized density functional theory(DFT) calculations were performed with Cambridge Serial Total Energy Package(CASTEP) codes [17,18]. The interactions between atomic core and valence electrons were described by ultrasoft pseudopotentials [19,20]. The electronic exchange-correlation energy was treated under local density approximation(LDA) [21]. The cut-off energy of the plane wave basis set was 400 eV for all of the cases, and 120 and 126 k-points were employed in the irreducible Brillouin zone for the cubic austenite and tetragonal distorted phases, respectively. The convergence tolerance for the calculations was selected as $10^{-6}$ eV/atom.

Table 1 Calculated structural, energy and magnetic parameters of the ground state of $Pt_{2-x}Mn_{1+x}Ga$ alloys

| Alloys | Structure | $a,b$(Å) | $c$(Å) | $c/a$ | $M_{tot}(\mu_B)$ | $M_{MnA}(\mu_B)$ | $M_{MnB}(\mu_B)$ | $\Delta E$(meV/atom) | $\Delta M(\mu_B)$ |
|---|---|---|---|---|---|---|---|---|---|
| $Pt_2MnGa$ | cubic | 6.09 | 6.09 | 1.00 | 3.65 | - | 3.51 | | |
| | tetragonal | 5.58 | 7.25 | 1.30 | 3.98 | - | 3.16 | -77.60 | 0.33 |
| $Pt_{1.75}Mn_{1.25}Ga$ | cubic | 6.04 | 6.04 | 1.00 | 2.43 | -3.42 | 3.28 | | |
| | tetragonal | 5.55 | 7.16 | 1.29 | 3.15 | -2.55 | 3.00 | -57.55 | 0.72 |
| $Pt_{1.5}Mn_{1.5}Ga$ | cubic | 5.99 | 5.99 | 1.00 | 1.88 | -3.30 | 3.07 | | |
| | tetragonal | 5.47 | 7.17 | 1.31 | 2.40 | -2.02 | 2.73 | -48.44 | 0.52 |
| $Pt_{1.25}Mn_{1.75}Ga$ | cubic | 5.94 | 5.94 | 1.00 | 1.08 | -3.02 | 2.90 | | |
| | tetragonal | 5.43 | 7.11 | 1.31 | 1.67 | -1.14 | 2.24 | -70.96 | 0.59 |
| $Mn_2PtGa$ | cubic | 5.88 | 5.88 | 1.00 | 0.56 | -2.63 | 2.73 | | |
| | tetragonal | 5.37 | 7.04 | 1.31 | 0.85 | -0.63 | 1.24 | -81.06 | 0.29 |

3. Results and discussion

$Pt_2MnGa$ crystallizes in $L2_1$ structure [22], in which Pt atoms occupy A(0.0, 0.0, 0.0) and C(0.5, 0.5, 0.5) sites, while Mn and Ga atoms occupy B(0.25, 0.25, 0.25) and D(0.75, 0.75, 0.75) sites, respectively. Similar to many $Mn_2$-based Heusler alloys, the cubic phase of $Mn_2PtGa$ crystallizes in $Hg_2CuTi$ structure [15], in which Mn atoms occupy A(0, 0, 0) and B(0.25, 0.25, 0.25) sites, while Pt and Ga atoms occupy C(0.5, 0.5, 0.5) and D(0.75, 0.75, 0.75) sites, respectively. Their atomic configurations are shown in Fig. 1. The $Pt_{2-x}Mn_{1+x}Ga$ alloys can be obtained by replacing Pt with Mn in $Pt_2MnGa$. According to the occupancy rule in Heusler compounds, the atoms with more valence electrons tend to occupy A and C sites, while the atoms with less valence electrons tend to occupy B and D sites. Since the valence electrons of Pt, Mn and Ga are $10(5d^96s^1)$, $7(3d^54s^2)$ and $3(4s^24p^1)$, respectively, the excess Mn atoms will replace the Pt atoms at A sites gradually in the process of the composition changing from $Pt_2MnGa$ to $Mn_2PtGa$. We label the excess and the original Mn atoms as $Mn_A$ and $Mn_B$, respectively.

In the first step, we have performed geometry optimization to determine the theoretical lattice parameter. The equilibrium lattice constant of the cubic $Pt_2MnGa$ is 6.09 Å at ground state(0 K), which is in good agreement with the experimental value of 6.16 Å at room temperature [22]. For $Mn_2PtGa$, the equilibrium lattice constant of the cubic phase is 5.88 Å. As illustrated in Table 1, with Mn content increasing, the lattice parameter of $Pt_{2-x}Mn_{1+x}Ga$ alloys in cubic phase decreases monotonically due to that the radius of Mn atom is smaller than that of Pt atom. We further find that $Pt_2MnGa$ is ferromagnetic while $Mn_2PtGa$, which is analogous to $Mn_2NiGa$ [13], is ferrimagnetic.

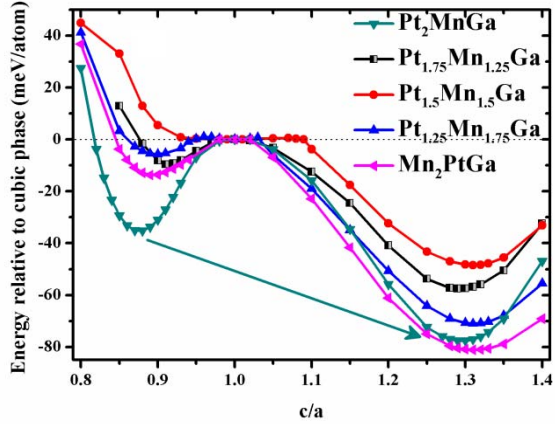

Fig. 2. The total energy differences between the tetragonal distorted and cubic austenite phases as a function of *c/a*. The zero point is corresponding to the cubic austenite phase.

Fig. 2 shows the total energy differences between the tetragonal distorted and cubic austenite phases calculated as a function of *c/a*: $\Delta E=E(c/a)–E(c/a=1)$. The zero point is corresponding to the cubic austenite phase. The corresponding magnetic moments and other physical parameters are summarized in Table 1. We can find that, besides $Pt_{1.5}Mn_{1.5}Ga$, there are two energy minimums for the energy curves: a local one at *c/a*<1, and a global one at *c/a*>1, which is energetically favorable. For $Pt_{1.5}Mn_{1.5}Ga$, there is only one energy minimum located at *c/a*=1.31. Based on experimental results, a system in which a martensitic transformation can occur should meet two requirements: (i) the thermodynamic driving force(*ΔE*) is large enough to overcome the resistance of phase transformation; (ii) the degree of distortion(*c/a*) is moderate, the value of which is always close to 1.20–1.30 [23]. As illustrated in Table 1, the driving force *ΔE* exhibits a distinct rule: it decreases firstly and then increase again with increasing Mn content. The driving force of $Pt_{1.5}Mn_{1.5}Ga$(-48.44 meV/atom) is the smallest. This value is considerably larger than the calculated driving force of $Ni_2MnGa$ [11], $Mn_2NiGa$ [13] and $Ni_2FeGa$ [11]. In addition, as listed in Table 1, for all the series, the *c/a* ratios are in the vicinity of 1.30. Thus, the martensitic transformation is prone to be available in $Pt_{2-x}Mn_{1+x}Ga$ alloys. The clarification of the martensitic transformation in $Pt_2MnGa$ in experiments [16] indicates the credibility of our calculations. Moreover, similar to the case in $Ni_{1.8}Pt_{0.2}MnGa$ [8], when allowing the tetragonal distorted phase corresponding to the energy minimum with c/a<1 to relax, a modulated structure has a great possibility to appear. In this case, as the arrow points out in Fig. 2, the transformation from the modulated structure to tetragonal phase(c/a>1) is likely to appear if a magnetic field is applied [8,24]. That is, a large MFIS is likely to appear in $Pt_{2-x}Mn_{1+x}Ga$(x=0, 0.25, 0.75, 1) alloys. This property has an important application on magnetic actuators.

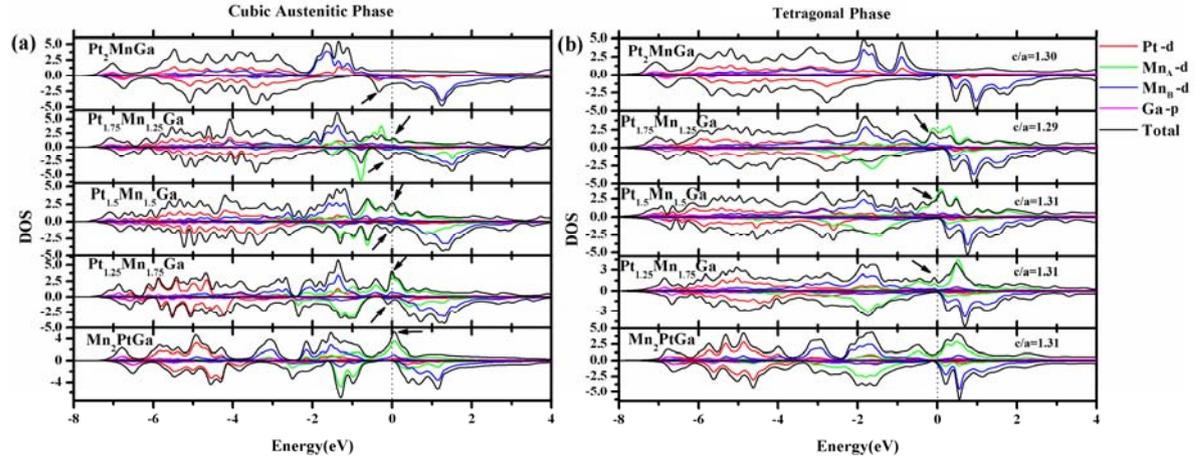

Fig. 3 Total and partial DOS of $Pt_{2-x}Mn_{1+x}Ga$ alloys in both cubic (a) and tetragonal phases (b).

It is known that the phase stability of the system is strongly influenced by the orbital hybridizations between the *p* state of the main-group elements and the *d* state of the transition metals [25]. In Heusler alloys, the strong *p-d* hybridization will give rise to strong covalent interactions and stabilize the structure. In addition, the phase stability will be sensitively influenced by the electronic structure around the Fermi level. It has been verified from theory and experiment in $Ni_2Mn_{1+x}Sn_{1-x}$ alloys that the sharp peak near the Fermi level plays an important role on the occurrence of the martensitic transformation in the ferromagnetic states [26]. The spin polarized total and partial density of state(DOS) of cubic austenite and the tetragonal phases are shown in Fig. 3. We can find that, there are sharp DOS peaks around Fermi level for all the cubic phases. Thus, these compounds in cubic phase are tending to distort to tetragonal phase to stabilize the system. By further analyzing the partial DOS, we can find that the DOS peak in the down-spin channel is mainly attributed to the hybridization between Pt-$5d$ and $Mn_B$-$3d$ states, while the DOS peak in the up-spin channel is mainly attributed to the $Mn_A$-$3d$ and $Mn_B$-$3d$ states. Upon tetragonal distortion, the peak in the down-spin channel in $Pt_2MnGa$ and that in the up-spin channel in $Mn_2PtGa$ shift through the Fermi level and a pseudogap forms at the Fermi level, resulting in the lowering of the total energy. For $Pt_{2-x}Mn_{1+x}Ga$(x=0.25, 0.5, 0.75) alloys, though the DOS peaks still exist around the Fermi level in the up-spin channel, the peaks in the down-spin channel also shift through Fermi Level and give rise to a pseudogap at the Fermi Level. This pseudogap formation at the Fermi Level also leads to the lowering of the total energy and stabilize the tetragonal phases [27]. However, the DOS peaks in the up-spin channel still give rise to a degree of instability to these tetragonal phases. Thus, the actual martensitic phase may be a modulated structure based on the tetragonal phase. Since our magnetic configurations are assumed to be collinear in this work, the instability of the martensitic phase may be also due to the non-collinearity of their actual magnetic structure. In addition, among these tetragonal phases, the DOS peak around the Fermi level in $Pt_{1.5}Mn_{1.5}Ga$ is the largest. This is the reason why the driving force of $Pt_{1.5}Mn_{1.5}Ga$ is the smallest.

The magnetic properties of the cubic austenite and tetragonal phases are also investigated in this work. The total magnetic moment per formula unit and atomic magnetic moment are listed in Table 1. It can be found that $Pt_2MnGa$ is ferromagnetic in both of the cubic austenite and tetragonal phases. For $Mn_2PtGa$, $Mn_A$ and $Mn_B$ are antiferromagnetically coupled in both phases, and the total magnetic moment is not zero due to that the magnetic moments of $Mn_A$ and $Mn_B$ are not equal. The total magnetic moment of the tetragonal phase for $Mn_2PtGa$ is 0.85 $\mu_B/f.u.$ This value is very close to the experimental value(0.80 $\mu_B/f.u.$ in 7 T at 1.9 K) obtained by Nayak et al. [15] For the couplings between $Mn_A$ and $Mn_B$ in both of the cubic austenite and tetragonal phases are antiferromagnetic, the total magnetic moments of $Pt_{2-x}Mn_{1+x}Ga$ alloys in both phases are decreasing with Mn content increasing. And similar to the case in $Mn_2PtIn$ [12], the total magnetic moment of the tetragonal phase is a little larger than that of the cubic austenite phase in all the series. The difference of the total magnetic moment between tetragonal and cubic austenite phases ($\Delta M$) are summarized in Table 1. Another feature of the change of the magnetic moment is that the variation quantity of the magnetic moment of $Mn_A$ across the tetragonal distortion is remarkably larger than that of $Mn_B$. This phenomenon is understandable. Since $Mn_A$ is the nearest neighbor of Ga while $Mn_B$ is the next nearest neighbor of Ga, the *p-d* hybridization of $Mn_A$-Ga is more sensitive to the structure distortion than that of $Mn_B$-Ga.

4. Conclusions

In conclusion, the electronic structure, magnetism and phase stability of Heusler-type $Pt_{2-x}Mn_{1+x}Ga$(x=0, 0.25, 0.5, 0.75, 1) alloys are investigated using the first-principles calculations. The magnetic martensitic transformation with a large thermodynamic driving force and a moderate strain energy is prone to occur in the all series. A large MFIS is likely to appear in $Pt_{2-x}Mn_{1+x}Ga$(x=0, 0.25, 0.75, 1) alloys. Electronic structure calculations reveal that the tetragonal phase is stabilized upon the distortion because of the pseudogap formation at the Fermi Level. The magnetic properties are also investigated and the total magnetic moment of the tetragonal phase is a little larger than that of the cubic austenite phase in all the series. Since the martensitic transformation temperature and the magnetic structure can be tuned by regulating the chemical composition, our work provides a guide to searching for new-type FSMAs based on Pt-Mn-Ga alloys.

**Acknowledgements**

This work is supported by the National Natural Science Foundation of China in Grant Nos. 51301119, 51301195 and 51171206, the Natural Science Foundation for Young Scientists of Shanxi in Grant No. 2013021010-1, Youth Foundation of Taiyuan University of Technology (No. 1205-04020102).

Captions:

Fig. 1. The atomic configuration of stoichiometric $Pt_2MnGa$ (a) and $Mn_2PtGa$ (b).

Fig. 2. The total energy differences between the tetragonal distorted and cubic austenite phases as a function of *c/a*. The zero point is corresponding to the cubic austenite phase.

Fig. 3 Total and partial DOS of $Pt_{2-x}Mn_{1+x}Ga$ alloys in both cubic (a) and tetragonal phases (b).